\documentclass[preprint,12pt]{elsarticle}

\usepackage{amssymb}
\usepackage{color,soul}

\journal{Nuclear Physics B}

\begin{document}

\begin{frontmatter}



\title{Generation of entanglement between quantum dot molecule with the presence of phonon effects in a voltage-controlled junction}


\author{Elaheh Afsaneh\corref{cor1}\fnref{L1}}
\ead{elahehafsaneh@gmail.com}

\author{Malek Bagheri Harouni\corref{cor1}\fnref{L1,L2}}
\ead{m-bagheri@phys.ui.ac.ir}

\address[L1]{Department of Physics, University of Isfahan, Hezar Jarib, Isfahan 81746, Iran}
\address[L2]{Quantum optics group, Department of Physics, University of Isfahan, Hezar Jarib, Isfahan 81746, Iran}

\cortext[cor1]{Corresponding author}



\begin{abstract}
We investigate the generation of entanglement through a quantum dot molecule under the influence of vibrational phonon modes in a bias voltage junction. The molecular quantum dot system is realized by coupled quantum dots inside a suspended carbon nanotube. We consider the dynamical entanglement as a function of bias voltage and temperature by taking into account the electron-phonon interaction. In order to generate the robust entanglement between quantum dots and preserve it to reach the maximal achievable amount steadily, we introduce an asymmetric coupling protocol and apply the easy tunable bias voltage-driven field. For an oscillating bias voltage, the time-varying entanglement can periodically reach the maximum revival.
In thermal entanglement dynamics, the phenomena of thermal entanglement degradation and thermal entanglement revival are observed which are intensively affected by the strength of phonon decoherence. The revival of entanglement shows a larger value for a higher phonon coupling.
\end{abstract}


\end{frontmatter}

\section{Introduction}
\label{Introduction}
Quantum entanglement has been considered as a crucial resource in quantum information processing which utilizes the non-local correlations without any classical analogs \cite{Peres,Horodecki,Nielsen,Ishizaka}.  
Generation and control of the entanglement have attracted a great deal of attention in different physics fields, such as atomic structures \cite{Pan,Khandekar}, photonic systems \cite{Zapletal,Sorelli}, and semiconductor quantum dots \cite{Rojas1,Rojas2,Keating}.
Quantum dots(QDs) as artificial atoms play prominent roles in quantum information studies where their discrete energy levels can be easily tuned by applying gate voltages \cite{Loss,Hanson-2}. 
Entanglement generation of an electron was studied by using a single-level quantum dot which was connected to one input and two output leads \cite{Oliver}. In this system, the process of electron entanglement formation was obtained as an analog to the entanglement of photons produced by a parametric amplifier. 

Also, quantum dot molecules(QDMs) formed by coupled QDs are taken into account as promising candidates for investigating and generating the quantum entanglement \cite{Bayer,Otten,Mendes}. 
The process of entanglement generation in double quantum dots(DQDs) is studied under the influence of the environment and is controlled by an external driving field \cite{Rojas1}. Particularly, the dynamical formation of entangled states in DQDs is explored through an external potential difference to demonstrate how this system can be controlled by an applied electric field \cite{Rojas2}. 
For realizing DQD systems, carbon nanotube (CNT) provides a potential setting \cite{Postma,Graber}. 
In this structure, a carbon nanotube bridges from one reservoir to another one, and allows electrons to be confined by applying gate voltage. This procedure provides constructing a DQD. 
A hybrid quantum system including a carbon nanotube double quantum dot and two nitrogen-vacancy(NV) centers is proposed for the generation of entanglement \cite{Song}.
In order to obtain steady-state entanglement, a carbon nanotube double quantum dot is considered as an environment of NV center. 
Furthermore, a suspended CNT which can oscillate freely is employed for constructing quantum dots \cite{Walter,Tang}.  
In this case, the suspension of CNT leads to the existence of vibrational phonons. These phonons play the role of the environment and have a profound effect on QDs \cite{Benyamini}. 

Real quantum systems due to contacting with surrounding environments, inevitably suffer from decoherence. These kinds of interactions lead to decreasing quantum correlations. 
For quantum entanglement correlations, loss of coherence is considered as a major obstacle which leads to entanglement degradation \cite{Yu}.  
Applying control techniques allows systems to suppress decoherence \cite{Yao,Khodjasteh}. 
On the other hand, there are some situations in which the interactions between the quantum system and environment can produce quantum correlations such as entanglement revival \cite{Zhang,Lu}. 
In other words, environments may operate as control elements for these cases \cite{Shi1}. 
The effect of the environment has a significant role in the generation and maximization of entanglement. For example, it is shown that due to the interaction of two qubits with a common heat bath and without any direct interaction, the entanglement can be created \cite{Braun}. 
Among various approaches considering the systems with an electron environment, the coupling of electron-phonon can be described through the quantum master equation in both Markovian \cite{Nazir,Rojas1,Rojas2} and non-Markovian \cite{Mogilevtsev,Reiter} regimes. 
Moreover, exploring the dynamics of quantum entanglement in a noisy environment shows that the coupling of a qubit with a non-Markovian environment can induce damping which resulted in the entanglement revival \cite{Shi1}. 
 
In the present contribution, we propose a molecular quantum dot system inside a suspended carbon nanotube with oscillatory phonon modes that is coupled with two metallic contacts. We study this structure to investigate the generation of steady entanglement and preserve it through a voltage-biased junction.
The dynamic of entanglement in the introduced QDM as an open quantum system is studied via a Markovian master equation under the effect of phonon decoherence. 
To engineer the influence of electron-phonon interaction on the entanglement evolution, we employ the strategy of asymmetric coupling for the coupled quantum dots. 
In addition, to control this correlation feature, the external bias voltage as a tunable device is applied in both types of constant and periodic time-varying fields. 
We first calculate the density matrix of our QDM system and then achieve the dynamical entanglement between the coupled quantum dots including the phonon influence. 
To characterize the entanglement of the QDs, we calculate the concurrence and explore its dependence on some physical parameters. 
Particularly, we study the dependence of concurrence on the external bias voltage and analyze the influence of temperature on the concurrence.
Through the time evolution of concurrence due to the effect of the applied oscillating bias voltage, the entanglement revival reveals periodically. 
The behavior of concurrence in response to varying temperature causes the phenomena of thermal entanglement degradation and revival. 

The outline of this paper is organized as follows: In Sec. \ref{Model}, we propose the model describing the quantum dot molecule with phonon modes in a bias voltage junction, and also describe the relevant phonon transformation. 
In Sec. \ref{Dynamics}, we derive the Markovian master equation and obtain the concurrence quantity by using the definition of the asymmetric factor for the present QDM setup. 
In Sec. \ref{Results}, we observe and discuss the results for the dynamics of concurrence against the bias voltage and temperature changes. Finally, we briefly conclude in Sec. \ref{Conclusion}. 
\section{Model} \label{Model}

\begin{figure}[h]
\centering
\includegraphics[scale=0.4]{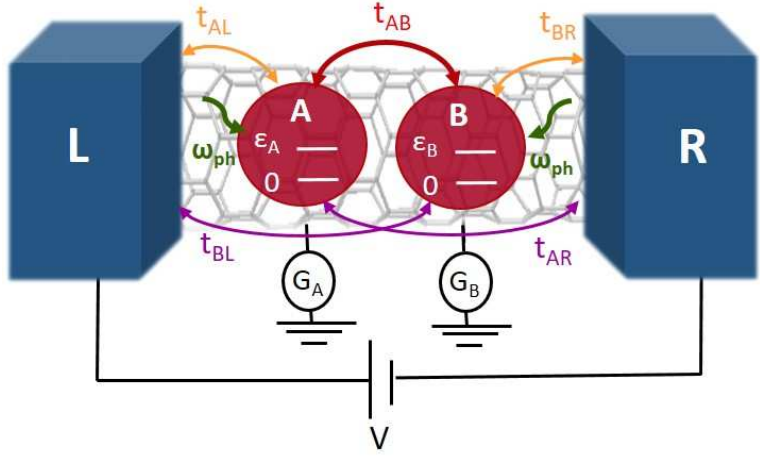}
\caption{Carbon nanotube double quantum dot molecule: A QDM system consists of quantum dots $A$ and $B$ are located in a suspended CNT.  Each QD is coupled to local phonon modes with frequency $\omega­_{ph}$ and is connected to the reservoirs $L$ and $R$ with tunneling amplitudes of $QD_{\alpha}$ to reservoir $\nu$ as $t_{AL}$, $t_{AR}$, $t_{BR}$, $t_{BL}$ . Two quantum dots are coupled together with the inter-dot tunneling amplitude $t_{AB}$. Energy levels of QDs, $\varepsilon_A$ and $\varepsilon_B$ are tuned by gate voltages $G_A$ and $G_B$, respectively. Reservoirs are held in the potential difference $V$.}\label{figure1}
\end{figure}

The physical system under study is a suspended carbon nanotube double quantum dot with vibrational phonon modes as an open quantum system in contact with voltage-biased reservoirs that is schematically shown in Fig.(\ref{figure1}). 

In the actual experiment, the energy states of a CNT with finite length are quantized. Therefore, the finite length CNT has energy levels like a quantum dot with discrete energy levels \cite{Buitelaar,Guinea}. 
Here, a finite length CNT is considered and connected with two metallic electrodes $L$ and $R$. 
To create a double quantum dot(DQD) within the carbon nanotube, a local gate is applied in the middle of CNT to build a center barrier. This tunnel barrier works as an inter-dot coupling, $t_{AB}$, that can separate two quantum dots $QD_A$ and $QD_B$(shown in Fig.(\ref{figure1})). 
In addition, the energy levels of each quantum dot can be controlled by individual gates $G_A$ and $G_B$. 
To have vibrational phonon modes in QDs, a suspended CNT is required. For this purpose, the center barrier of CNT is kept fixed while its two lateral sides are allowed to oscillate freely using the etching technique \cite{Huttel-1}. In this case, the suspended lateral parts of CNT located between the center tunnel barrier and metallic contacts would be considered as quantum dots with vibrational phonons \cite{Walter,Tang}.
These characteristics of quantum dot molecule are introduced by the Hamiltonian of quantum dot molecule $H_{QDM}$.
Also, in figure (\ref{figure1}) the normal metal reservoirs $L$ and $R$ are held at chemical potentials $\mu_L$ and $\mu _R$($\mu_L>\mu_R$) to provide the bias voltage $V =\mu_L-\mu_R$.

The total Hamiltonian for the proposed system is given by:
\begin{equation}\label{Htot}
H=H_{QDM}+H_{Res}+H_{int}.
\end{equation}
To define the Hamiltonian of quantum dot molecule $H_{QDM}$, we use the Anderson-Holstein model \cite{Holstein,Mahan} as:
\begin{equation}\label{HQDM}
H_{QDM}=\sum_{\alpha=A,B}\left(\varepsilon_{\alpha}d^{\dagger}_{\alpha}d_{\alpha}
+\omega_{ph}b^{\dagger}_{\alpha}b_{\alpha}+t_{ph}\hat{n}_{\alpha}(b^{\dagger}_{\alpha}+b_{\alpha})\right)+t_{AB}(d^{\dagger}_{A}d_{B}+d_{A}d^{\dagger}_{B}).
\end{equation}
Here, $H_{QDM}$ consists of the Hamiltonian of quantum dots $H_{dots}=\sum_{\alpha=A,B}\varepsilon_{\alpha}d^{\dagger}_{\alpha}d_{\alpha}+ t_{AB}(d^{\dagger}_{A}d_{B}+d_{A}d^{\dagger}_{B})$, 
Hamiltonian of phonons $H_{ph}=\sum_{\alpha=A,B}\omega_{ph}b^{\dagger}_{\alpha}b_{\alpha}$ and the electron-phonon interaction Hamiltonian $H_{el-ph}=\sum_{\alpha=A,B}t_{ph}\hat{n}_{\alpha}(b^{\dagger}_{\alpha}+b_{\alpha})$. In Eq.(\ref{HQDM}), $\varepsilon_{\alpha}$ denotes the electronic energy level of quantum dot 
$\alpha$, $\omega_{ph}$ is the local phonon frequency, $t_{ph}$ shows the strength of the electron-phonon coupling, and $t_{AB}$ is the inter-dot hopping amplitude which 
with no loss of generality is taken as a real parameter. Moreover, $d^{\dagger}_{\alpha}(b^{\dagger}_{\alpha})$ indicates the electron(phonon) creation operator, and 
$\hat{n}_{\alpha}=d^{\dagger}_{\alpha}d_{\alpha}$ is the occupation operator for quantum dot $\alpha=A,B$. It should be noted that the fermionic operators $d_{\alpha}$ fulfills the fermionic anti-commutator whereas $b_{\alpha}$ is bosonic in nature.
The Hamiltonian of reservoirs can be written as:
\begin{equation}
H_{Res}=\sum_{k, \nu=L,R} \epsilon_{k\nu} {c}^{\dagger}_{k\nu}c_{k\nu},
\end{equation}
which contains the non-interacting electrons where $\epsilon_{k\nu}$ denotes the energy level of reservoir $\nu=L,R$ and ${c}^{\dagger}_{k\nu}$ creates an electron with momentum $k$ in lead $\nu$. Here, reservoirs are assumed completely spin-polarized in which the spin of electrons can not be distinguished.
The interaction Hamiltonian in Eq.(\ref{Htot}) corresponds to tunneling between QDs and electrodes which is described as:
\begin{equation}
H_{int}=\sum_{k,\alpha,\nu} \left(t_{\alpha \nu}c^{\dagger}_{k\nu}d_{\alpha}+t^{*}_{\alpha \nu} c_{k\nu}d^{\dagger}_{\alpha}\right).
\end{equation}
In which, $t_{\alpha \nu}$ denotes the tunneling amplitude between $QD_{\alpha}$ and reservoir $\nu$ which is assumed energy and momentum independent.
The tunneling rates of the reservoir $\nu$ is characterized by $\Gamma_{\alpha \nu}=2\pi N^0_{\nu}|t_{\alpha \nu}|^2$. This parameter can be defined in the wide-band limit (WBL) \cite{Dong} where, both parameters $ N^0_{\nu}$, the density of states of the lead $\nu$ and $ t_{\alpha \nu}$ are assumed constant without energy dependency. This assumption allows us to have the tunneling rates as an energy-independent feature. Here, we consider the total tunnel-coupling strength for our QDM system as $\Gamma =\sum_{\alpha,\nu}\Gamma_{\alpha \nu}$. Moreover, for weak tunneling to the electronic reservoirs, the tunneling rates assumed as the smallest energy scale in this system, $\Gamma_{\alpha \nu} \ll k_{B}T$. This condition provides the lowest order of tunnel-coupling and also, it allows the reservoirs to stay in thermal equilibrium. 

In order to proceed further, the $H_{QDM}$ in Eq.(\ref{HQDM}) can be diagonalized by applying the  polaron Lang-Firsov transformation which is given through the following relation \cite{Mahan,Lang}:
\begin{equation}\label{TransformedHQDM}
\bar{H}=e^{S}He^{-S}, \quad S=\sum_{\alpha} g_{ph}\hat{n}_{\alpha}(b^{\dagger}_{\alpha}-b_{\alpha}).
\end{equation}
Here, $g_{ph}=\frac{t_{ph}}{\hbar \omega_{ph}}$ denotes the coupling of the local phonon modes with energy $\hbar \omega_{ph}$. 
Since the phonons with the small energy scale are considered as acoustic phonons \cite{Huttel-2}, so the phonons in our QDM system with the energy range of a few meV treated as acoustic modes. Also, it is reported that in semiconducting systems, the longitudinal modes are more dominant for acoustic phonons \cite{Pennington}. Therefore, in the present suspended CNT-DQD system, we deal with the phonons in the type of longitudinal acoustic modes.

The applied transformation eliminates the electron-phonon coupling term in $H_{QDM}$(Eq.(\ref{HQDM})). So, the transformed $QDM$ Hamiltonian is obtained as:
\begin{equation}\label{TransformedHQDM}
\bar{H}_{QDM}=\sum_{\alpha}\left( \bar{\varepsilon}_{\alpha}d^{\dagger}_{\alpha}d_{\alpha}
+\hbar\omega_{ph}b^{\dagger}_{\alpha}b_{\alpha}\right) +t_{AB}\left(X^{\dagger}_{A}d^{\dagger}_{A}X_{B}d_{B}+H.c.\right),
\end{equation}
where H.c. represents the Hermitian conjugate, $X_{\alpha}=e^{g_{ph}(b_{\alpha}-b^{\dagger}_{\alpha})}$ shows the polaron operator and $ \bar{\varepsilon}_{\alpha}=\varepsilon_{\alpha}-g^{2}_{ph}\omega_{ph}$  indicates the renormalized QD energy levels. As a consequence of transformation introduced in 
Eq.(\ref{TransformedHQDM}), the Hamiltonian of the uncoupled reservoirs remains unchanged while the tunneling Hamiltonian is transformed as bellow:
\begin{equation}\label{TransformedHint}
\bar{H}_{int}=\sum_{k,\alpha,\nu} \left(t_{\alpha \nu}c^{\dagger}_{k\nu}X_{\alpha}d_{\alpha} +t^{*}_{\alpha \nu} c_{k\nu} X^{\dagger}_{\alpha}d^{\dagger}_{\alpha}\right).
\end{equation}
To study the dynamics of the system governs by the transformed Hamiltonian, in the next section we first calculate the quantum master equation(QME). Then, we investigate the dynamics of concurrence as a measure of entanglement by using the asymmetric factor definition.  
\section{Dynamics}\label{Dynamics}
To generate the steady entanglement in the present molecular quantum dot system and investigate the dynamics of this quantum correlation, we start from the Liouville-von Neumann equation of the whole system in the interaction picture \cite{Breuer}. 
The complete system consists of a central double quantum dot, electronic reservoirs, and oscillating phonon bath systems with Hilbert spaces $\mathcal{H}_{dots}$, $\mathcal{H}_{res}$, and $\mathcal{H}_{ph}$ respectively. 
Consequently, the total Hilbert space of the whole system is defined as $\mathcal{H}_{tot}=\mathcal{H}_{dots}\otimes \mathcal{H}_{res}\otimes \mathcal{H}_{ph}$. 
For suspended carbon nanotubes, it is reported that the electron-phonon interaction is a strong coupling \cite{Biercuk}. Therefore, we suppose that the present system is considered in the high enough phonon frequency to obtain the strong electron-phonon coupling. Also, for the assumption of weakly coupling of quantum dots with reservoirs, we can apply the Born-Markov approximation.
Therefore, the density matrix of the total system is approximately characterized by $\rho_{tot}(t) \approx \rho_{dots}(t)\otimes\rho_{res}(t)\otimes\rho_{ph}(t)$ for 
thermal equilibrium baths. By tracing out the bath degrees of freedom for both electronic reservoirs and phonon baths, the quantum master equation of the central QDM system is obtained as:
\begin{eqnarray}\label{QME}
\frac{d\rho_{s}(t)}{dt}&=&-\frac{i}{\hbar} [\bar{H}_{dots}(t),\rho_{s}(t)] \nonumber \\
&-& \frac{1}{\hbar^2}\int_{0}^{\infty}dt^{'}Tr_{res}Tr_{ph}\left[ \bar{H}_{int}(t),\left[\bar{H}_{int}(t-t^{'}),\rho_{tot}(t)\right] \right].\nonumber \\
\end{eqnarray}
The first term corresponds to the coherent evolution of the system while the second one raises due to the different sources of dissipation. We ignore the first term in the condition of  $t_{AB}\leq t_{\alpha \nu}\leq t_{ph}$ which means that electron-phonon coupling is larger than the coupling of dots with reservoirs, so the coherent dynamics of dots is negligible in the presence  of phonon interaction.
 By substituting the transformed Hamiltonians, $\bar{H}_{QDM}$ and $\bar{H}_{int}$ into Eq.(\ref{QME}), the reduced density matrix of QDM subsystem fulfills the following equation:
\begin{eqnarray}
\frac{d\rho_{s}}{dt}&=&\sum_{\alpha,\nu} (M^{+}_{\alpha \nu}\left[2d_{\alpha}\rho_{s}(t)d^{\dagger}_{\alpha}-d^{\dagger}_{\alpha}d_{\alpha}\rho_{s}(t)-\rho_{s}(t)d^{\dagger}_{\alpha}d_{\alpha} \right]\nonumber \\
&+&M^{-}_{\alpha \nu}\left[ 2d^{\dagger}_{\alpha}\rho_{s}(t)d_{\alpha}-d_{\alpha}d^{\dagger}_{\alpha}\rho_{s}(t)-\rho_{s}(t)d_{\alpha}d^{\dagger}_{\alpha} \right]).
\end{eqnarray}
In this equation, $M^{\pm}_{\alpha \nu}$ describe the tunneling rates as:
\begin{equation}\label{M+}
M^{+}_{\alpha \nu}= \Gamma_{\alpha \nu} \int d\omega \langle c_{\nu}c^{\dagger}_{\nu} \rangle \langle X^{\dagger}_{\alpha}X_{\alpha}\rangle 
=\Gamma_{\alpha \nu} \int d\omega [1-f_{\nu}(\omega+\epsilon_{\nu})]G^{+}_{\alpha}(\omega),
\end{equation}
and
\begin{equation}\label{M-}
M^{-}_{\alpha \nu}=\Gamma_{\alpha \nu} \int d\omega \langle c^{\dagger}_{\nu}c_{\nu} \rangle \langle X_{\alpha}X^{\dagger}_{\alpha}\rangle
=\Gamma_{\alpha \nu} \int d\omega f_{\nu}(\omega+\epsilon_{\nu})G^{-}_{\alpha}(\omega).
\end{equation}
In equations (\ref{M+}) and (\ref{M-}), the correlation function of reservoirs is defined as $\langle c_{\nu}c^{\dagger}_{\nu} \rangle=Tr_{res}[\rho_{res}c_{\nu}c^{\dagger}_{\nu}]$  which gives us the Fermi distribution function of lead $\nu$ by $f_{\nu}(\omega+\epsilon_{\nu})=\frac{1}{e^{\beta_e(\omega+\epsilon_{\nu})}+1}$. The temperature of electrons in both reservoirs is assumed the same and is shown by $T_e$ which provides $\beta_e=\frac{1}{k_BT_e}$.
The phonon-assisted correlation functions are demonstrated by $G^{\pm}_{\alpha}(\omega)$. Here, the negative correlation function is given by $G^{-}_{\alpha}(t)=Tr_{ph}[\rho_{ph}X_{\alpha}(t)X^{\dagger}_{\alpha}]$ \cite{Mahan}, and its Fourier transform which is $G^{-}_{\alpha}(\omega)=\int^{\infty}_{0} dt e^{i \omega t}G^{-}_{\alpha}(t)$ can be calculated as:
\begin{equation}
G^{-}_{\alpha}(\omega)=e^{-g(2N_{ph}+1)}\sum^{\infty}_{l=-\infty}I_{l} \left( 2g_{ph}\sqrt{N_{ph}(1+N_{ph})} \right)
\times  e^{l\frac{\omega_{ph}}{2}\beta}2\pi \delta(\omega-\omega_{ph}l).
\end{equation}
Here, the Bose-Einstein distribution function of phonon is $N_{ph}=\frac{1}{e^{\beta_{ph}\omega_{ph}}-1}$ with $\beta_{ph}=\frac{1}{k_BT_{ph}}$. 

Since phonons inside the CNT are close to reservoirs, phonon temperature is assumed equal to the electron temperature of electrodes, $T_{ph}=T_e=T$.
$I_l(z)$ denotes the modified Bessel function of the first kind of order $l$. The positive and negative correlation functions are related to each other by the relation $G^{+}_{\alpha}(\omega)=G^{-}_{\alpha}(-\omega)$.
Now with the achieved phonon-assisted correlation function of the system, we can proceed further to calculate concurrence as an entanglement measure in our QDM system.
\subsection{Concurrence}\label{Concurrence}
To quantify entanglement between two qubits, Wootters proposed the measure of concurrence for both pure and mixed states \cite{concurrence-1,concurrence-2}. 
In condensed matter systems, the entanglement of fermions is evaluated by fermionic concurrence \cite{Schliemann-2001,Schliemann-2001-2,Schliemann-2002,Majtey,Afsaneh2} in analog with Wootter's formula.  
Therefore, the entanglement between quantum dots including electrons as indistinguishable fermions can be calculated in the formalism of fermionic concurrence that we discussed in the previous paper \cite{Afsaneh2}. 
Studying the concurrence of quantum dot qubits has attracted considerable attention \cite{Nori-2} and particularly, this measure was determined for double quantum dot systems \cite{Rojas1,Rojas2,Afsaneh2}.  
Here, for our carbon nanotube quantum dot molecule, we define the concurrence as:
\begin{equation}\label{concurrence-eq1}
C(\rho)=Max [0,\lambda_1-\lambda_2-\lambda_3-\lambda_4],
\end{equation}
where, $\lambda_i,(i=1,2,3,4)$ are the non-negative eigenvalues of matrix $R$ in decreasing order, $\lambda_1>\lambda_2>\lambda_3>\lambda_4$. Matrix $R$ is evaluated as $R=\sqrt{\sqrt{\rho}\tilde{\rho}\sqrt{\rho}}$ with the density matrix of the system, $\rho $, and $\tilde{\rho}=(\sigma_y \otimes \sigma_y) \rho ^*(\sigma_y \otimes \sigma_y)$.  Here, $\rho^*$ shows the complex conjugate of the density matrix  and $\sigma_y$ represents the $y$ element of Pauli matrices.
The concurrence takes value in the interval between zero for the separable states and one unit magnitude for the maximally entangled states.

The basis states of the present QDM system with quantum dots $A$ and $B$(Fig.(\ref{figure1})) can be selected as $|\psi\rangle_{AB}=|\Phi\rangle_{A} \otimes |\Phi\rangle_{B}$. In which, $|\Phi\rangle_{A}$ and $|\Phi\rangle_{B}$ denote the states of quantum dot $A$ and $B$, respectively. We suppose that each dot as a single energy level contains unoccupied $|0\rangle_{\alpha} $ and occupied $|1\rangle_{\alpha} $ states with the energies $0$ and $\varepsilon_{\alpha}$($\alpha=A, B$), respectively. In other words, double occupancy is forbidden. Therefore, we can present the total form of the occupation states as $|0_{A},1_{A},0_{B},1_{B}\rangle=|0_{A},1_{A}\rangle\otimes |0_{B},1_{B}\rangle $.
Here, to obtain the concurrence of the present QDM system, we assume that QDs are not entangled initially and we consider the initial state of two coupled unentangled QDs as:
\begin{equation}\label{initial-asymmetric}
\rho(0)= \left[ {\begin{array}{cccc}
   1 & 0 & 0 & 0 \\
    0 & 0 & 0 & 0  \\
    0 & 0 & 0 & 0 \\
    0 & 0 & 0 & 0 \\
  \end{array} } \right].
\end{equation}
In the following, we define an asymmetric factor which quantifies the coupling of QDs asymmetrically and leads to providing the maximum achievable entanglement. 
\subsection{Asymmetric Factor}
Phonon decoherence in quantum systems can destroy the quantum correlation and causes entanglement degradation.    
One technique that can be used to reduce the environmental dissipation in the quantum dot system is employing the coupling of quantum dots to reservoirs asymmetrically\cite{Rojas1}.    
Recently, we showed that by using quantum dots with asymmetric coupling coefficients, robust entanglement is achieved for a QDM system in a Josephson junction \cite{Afsaneh2}. 
Here, to generate entanglement with a remarkable value and preserve it steadily, we define an asymmetric factor in which the coupling coefficients of QDs-metal leads are determined unequal. 
In this case, we assume that each quantum dot is coupled to the left and right reservoirs with different strengths that can be tuned by applying gate voltages.    
For this proposed structure, there is a quantum dot in the middle of quantum dot-reservoir coupling.   
To produce these kinds of setups, it was reported that the partial capacitors were used in a parallel double quantum dot system \cite{Hofmann}. In the coupled quantum dot systems, the partial capacitors mean that the tunneling barriers  should not be built with the same heights for creating quantum dots. It provides the relevant capacitors locating not absolutely in the parallel plane.  
In this case, the interaction between one dot and reservoir is assisted with an intermediate dot. 
This shows that by arranging the partial capacitors, the structure of our proposed system can be fabricated. 
  Therefore, we suppose that QDs can be connected with both near and far reservoirs with various non-zero couplings.  Then, the tunnel-coupling strength for each reservoir would be written as $T_{\nu}=\frac{T_{A\nu}+T_{B\nu}}{2} $.

 In which $T_{A\nu}(T_{B\nu})$ is the coupling strength of $QD_{A}(QD_{B})$ with the reservoir $\nu$. It means that our system needs four tunnel-coupling coefficients of $ T_{AL}$, $ T_{BL}$, $ T_{AR}$, and $ T_{BR}$ are shown in Fig.(\ref{figure1}). 
We introduce the asymmetric factor as \cite{Afsaneh2}:
\begin{equation}\label{asymmetric factor}
\kappa=\frac{\kappa_{A}+\kappa_{B}}{2},
\end{equation}
 where $\kappa^2_{\alpha}=|\frac{T_{\alpha L}-T_{\alpha R}}{T_{\alpha L}+T_{\alpha R}}|^2$. 
The amount of asymmetric factor ranges from zero for symmetric coupling coefficients, to one for the complete asymmetric situation.

The symmetric structure is defined when each QD is coupled to the left and right leads with the same coupling coefficient, $T_{\alpha L}=T_{\alpha R}$. This situation provides a minimum magnitude of asymmetric factor, $\kappa=0$. The asymmetric structure refers to the left-right different coupling coefficients, $T_{\alpha L} \neq T_{\alpha R}$, with $0<\kappa\leq1$ amount. Completely asymmetric configuration with $\kappa \simeq 1$ can be realized for the specific physical properties when the strength of coupling for the near-lead is much larger than the far-lead. 
This means that we have $T_{AL} \gg T_{AR}$($T_{BR} \gg T_{BL}$) for $QD_A$($QD_B$). 

In the next section, with the calculated density matrix and considering the required asymmetry factor, we present the results for the behavior of entanglement dynamics.
\section{Results}\label{Results}
In this section, we demonstrate the results of investigating dynamical entanglement for our QDM system against the bias voltage changes, and temperature-varying in Concurrence-Voltage and Concurrence-Temperature parts, respectively. 
For simplicity and with no loss of generality, we assume the frequency of phonons for each QD is the same, $\omega_{ph}$. 
In all results, the order of energy is supposed $meV$, and also the order of temperature is assumed $mK$. 
\subsection{Concurrence-Voltage}
For this QDM system, the effect of bias voltage on the concurrence is originated from the reservoir distribution function in the dissipation expression $\Gamma$. When the reservoirs are derived asymmetrically by the external bias voltage, their chemical potentials are changed as $\mu_{L}=\mu_{0}+eV$ and $\mu_{R}=\mu_{0}$. Therefore, the concurrence is affected by the bias voltage indirectly.
To explore the behavior of concurrence as a function of bias voltage, we study the concurrence-voltage curves(C-V characteristic) when the bias voltage is constant($dc$) as well as time-dependent($ac$). 
\subsubsection{Concurrence-$dc$ Voltage}
\begin{figure}[t]
\centering
\includegraphics[width=0.48\columnwidth]{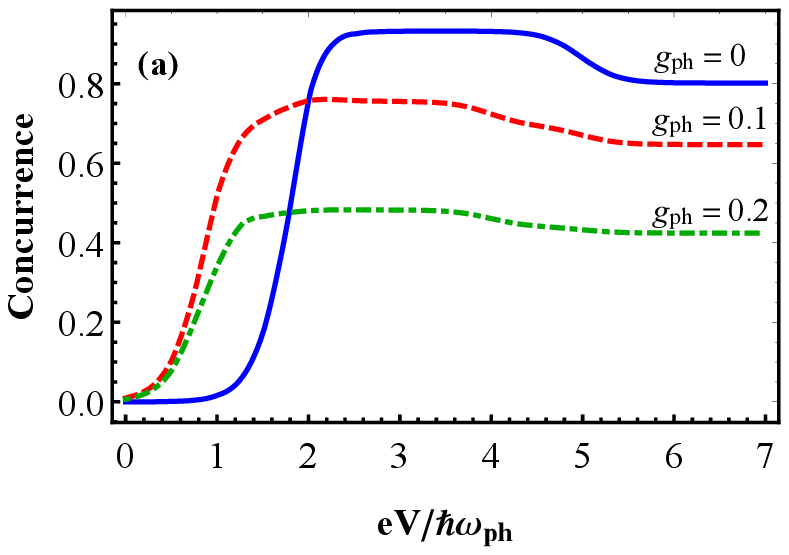}
\hspace{0.1cm}
\includegraphics[width=0.48\columnwidth]{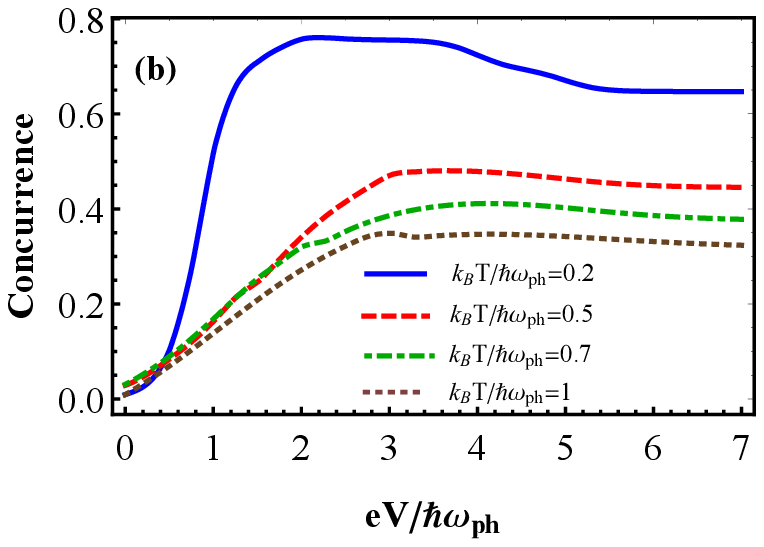}
\caption{Concurrence-Voltage behavior for constant bias voltage driving with asymmetric factor, $\kappa=0.55$ when (a): in certain phonon coupling strengths with fixed $\frac{k_{B}T}{\hbar\omega_{ph}}=0.2$. (b): in certain temperatures with fixed phonon coupling strength $g_{ph}=0.1$. The energy levels of quantum dots $A$ and $B$ are selected $\varepsilon_A / \hbar\omega_{ph}=2$ and $\varepsilon_B / \hbar\omega_{ph}=5$ respectively. }\label{Fig2}
\end{figure}

To observe the concurrence dependency on the constant bias voltage, we illustrate C-V characteristic curve for asymmetrically coupled quantum dots with asymmetric factor, $\kappa=0.55$ in figure(\ref{Fig2}).  
Panel $a$ of this figure demonstrates the influence of electron-phonon coupling strength on the concurrence.
In very low bias voltage($\varepsilon/ \hbar\omega_{ph}<1$), quantum dots are not completely excited and consequently the measure of entanglement is very small. By increasing voltage, concurrence shows increment until $\varepsilon/ \hbar\omega_{ph}=2$. 
Refer to the energy levels of quantum dots $A$ and $B$(assumed $\varepsilon_A / \hbar\omega_{ph}=2$ and $\varepsilon_B / \hbar\omega_{ph}=5$ respectively), concurrence in resonant with quantum dots energy levels shows behavior-changing which is similar to steps through the increment of bias voltage. 
All curves of Fig.(2)a are under the effect of resonant with quantum dots energy levels and exhibit changes in concurrence behavior. 
The main difference between these curves is the presence of phonon coupling. For the solid line with no phonon coupling, the steps due to the resonance are in complete shape while for the dashed and dot-dashed lines with the phonon coupling, the steps become smoother. With further increasing bias voltage, concurrence behaves steadily. This leads to observe the steady-state entanglement for high bias voltage.

In panel (b) of Fig.(\ref{Fig2}), we plot the C-V curve for a fixed phonon coupling $g_{ph}=0.1$ in some certain temperatures. This figure shows that although the phonon coupling is fixed for all C-V curves, concurrence observes more decrement in larger temperatures. In other words, due to the increase of temperature, thermal decoherence is raised which causes more concurrence decline.
 
Two panels of Fig.(\ref{Fig2}) express that by increasing the electron-phonon coupling strength in panel $a$ and raising the temperature in panel $b$, the entanglement shows degradation because of phonon and thermal decoherences respectively. 
An interesting point in this figure is that despite these two dissipative features, the concurrence amount is kept at a significantly steady amount. The main reason for this behavior of the system in preserving the entanglement is due to applying the coupling coefficients asymmetrically with non-zero asymmetric factors. 
\subsubsection{Concurrence-$ac$ Voltage}
To further investigate the parameters which affect the entanglement of the QDM setup, we apply time-dependent voltage. 
Usually, the time evolution of entanglement under the periodically driven voltage is studied for structures with multi subsystems \cite{Kim,Ponte}. Here, we study the dynamics of entanglement for our bipartite system when it is driven periodically with an ac periodic voltage as:
\begin{equation}
V_{ac}(t)=V^{0}_{dc}+V^{0}_{ac}Cos(\omega_{ac}t),
\end{equation}
where $V_{dc}$ indicates the amplitude of constant voltage while $V^{0}_{ac}$ and $\omega_{ac}$ denote the amplitude and frequency of oscillating voltage, respectively.
Applying a time-dependent voltage to the system leads to having some parameters depending on time such as the energy level of QDs, the electrochemical potential of reservoirs, and the tunnel-coupling coefficients.
As mentioned before, we suppose our system in the WBL regime so the tunnel-coupling parameters are assumed constant and energy-independent.

\begin{figure}[t]
\centering
\includegraphics[width=0.48\columnwidth]{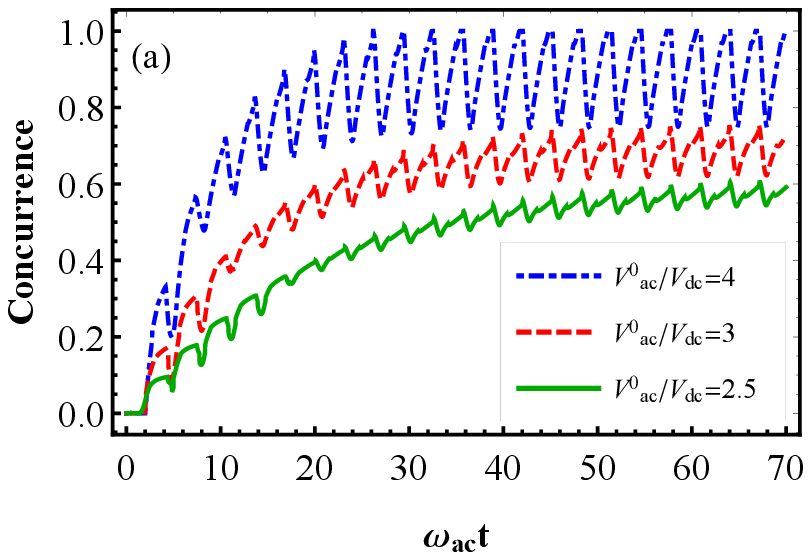}
\hspace{0.1cm}
\includegraphics[width=0.48\columnwidth]{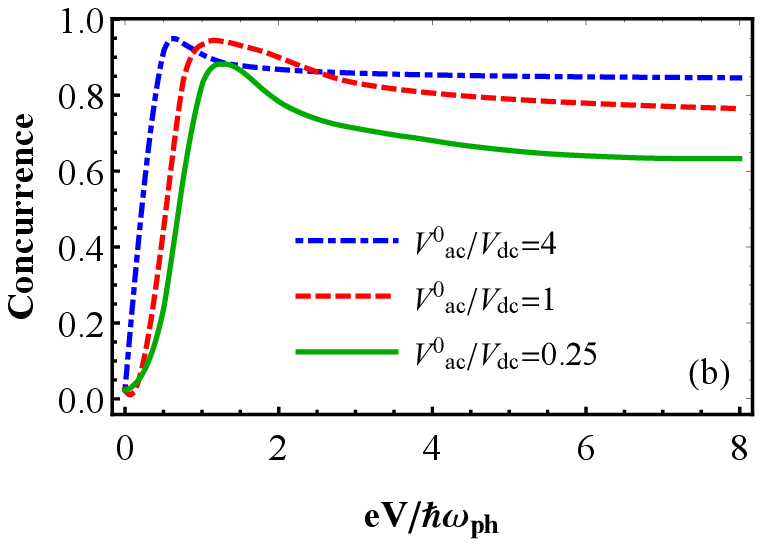}
\caption{Concurrence evolution for periodic bias voltage $V_{ac}(t)=V_{dc}+V_{ac}^0 Cos(\omega_{ac}t) $ with $\frac{eV_{dc}}{\hbar \omega_{ph}}=1$, $\frac{\omega_{ac}}{\omega_{ph}}=2$, $g_{ph}=0.1$, $\kappa=0.58$ and $\frac{k_{B}T}{\hbar \omega_{ph}}=0.2$ for (a): time evolution of concurrence (b): average concurrence against periodic bias voltage}\label{Fig3}
\end{figure}

In figure(\ref{Fig3}), we show the dynamics of concurrence for a fixed phonon coupling $g_{ph}=0.1$ under the harmonic bias voltage in two panels (a) and (b). In panel (a), the time evolution of concurrence exhibits periodic time dependence. 
The oscillating pattern of concurrence displays increasing behavior for each driving cycle and it revives with a higher magnitude. As time elapses enough($t\rightarrow \infty$), it reaches the stationary periodic manner. 
Here, the periodic revival concurrence strongly depends on the amount of bias voltage amplitude. Such that for higher bias voltage amplitude, the shape of revival for each cycle tends to verify from the cosine oscillation shape. However, in this case, the revival concurrence can achieve higher amounts due to the larger driving voltage amplitudes. 
This leads to the maximum entanglement for $\frac{V^{0}_{ac}}{V_{dc}}=4$ with an oscillatory behavior.

To observe the C-V characteristic against time-dependent voltage, we show the behavior of the time-averaged concurrence as a function of bias voltage in Fig.(\ref{Fig3})b. 
In this figure, the average concurrence increases more quickly with higher maximum amounts for more driving bias voltage amplitudes $V^0_{ac}/V_{dc}$. 
This behavior originates from the treatment of electrons with respect to the periodic bias voltage. In other words, due to the increase of voltage, electrons can oscillate faster which leads to raising the magnitude of average entanglement.
Furthermore, by applying larger time-dependent driving fields, the average concurrence reaches the stationary state with a robust amount.  
Totally, figure (\ref{Fig3}) demonstrates that the introduced QDM system can protect the entanglement between quantum dots under the control of time-varying bias voltage despite the presence of phonon decoherence.  
\subsection{Concurrence-Temperature}
\begin{figure}[t] 
\centering 
\includegraphics[width=0.48\columnwidth]{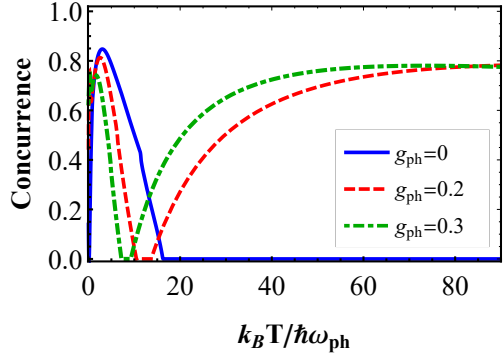} 
\caption{Concurrence-Temperature behavior for $\frac{eV}{\hbar\omega_{ph}}=0.1$ and $\kappa=0.63$.}\label{Fig4} 
\end{figure}

To obtain more information during the evolution of entanglement in the present system, we evaluate the behavior of the concurrence with respect to the temperature. 
The temperature-dependent nature of concurrence is originated from the temperature dependency of phonon Bose-Einstein distribution function $N_{ph}$ and fermionic distribution function of leads $f_{\nu}$. 
To observe the behavior of entanglement against temperature-varying, we plot the concurrence-temperature curve(C-T characteristic) which is shown in Fig.(\ref{Fig4}). 

This figure illustrates that in the beginning, due to the constant driven bias voltage, the unentangled electrons in the initial ground states are excited to the higher levels. 
After that, electrons find more opportunities to be entangled through the increase of temperature. This leads to an entanglement increment toward the maximum achievable value.  
Increasing more temperature causes the separation of the entangled electrons which is observed as a decreasing manner of concurrence. 
This decline behavior continues through a finite temperature interval to reach zero amount. At this point, entanglement disappears completely which means that all electrons are unentangled. 

Here, we call the complete disappearance of entanglement against temperature a thermal entanglement degradation(TED). 
In TED parts of figure(\ref{Fig4}), it is obvious that for larger phonon coupling strengths, concurrence reaches the lower maximum value and then collapses with a shorter TED interval. 
This kind of entanglement degradation was reported as a significant property for quantum dot systems \cite{Zou}. 
  
Applying more temperature provides electrons to move faster and consequently find more possibilities to be entangled again. Fig.(\ref{Fig4}) illustrates that the entanglement can reappear after a period of temperature absence.   
We name the phenomenon of entanglement rebirth after the complete disappearance versus temperature as a thermal entanglement revival (TER). 
In Fig.(\ref{Fig4}) for TER parts, it is shown that concurrence can be revived with a larger magnitude for a stronger phonon coupling coefficient.

After the revival, entanglement raises to reach a robust value and steadily continues through the temperature increment.   
In this figure, for C-T cure without any phonon coupling $g_{ph}=0$(solid line), the entanglement can not be revived after its disappearance.  
This behavior of entanglement reveals that the key reason for the revival phenomenon in the present system would be originated from the decoherence of surrounding phonons. 
In semiconductor systems, electron-phonon interaction is considered as a dominant term over the other relevant correlations \cite{Perebeinos}. 
Moreover, researches show that phonon interaction is strongly affected by the value of temperature \cite{Clear1}. It means that by increasing the phonon strength, the absorption of this parameter from the thermal bath can be improved.   
Therefore, we believe that in our QDM system, the electron-phonon effect plays a crucial role in the generation of the thermal entanglement revival. Also, the influence of this element is so strong that the steady-state TER is obtained with a robust amount for the higher temperature.  
 
To demonstrate the importance of phonon effect on the behavior of entanglement in the larger value of temperature, we investigate the temperature dependency of concurrence on the high-temperature(HT) approximation in the following. 
\subsubsection{High-Temperature Approximation} 
\begin{figure}[t]
\centering
\includegraphics[width=0.48\columnwidth]{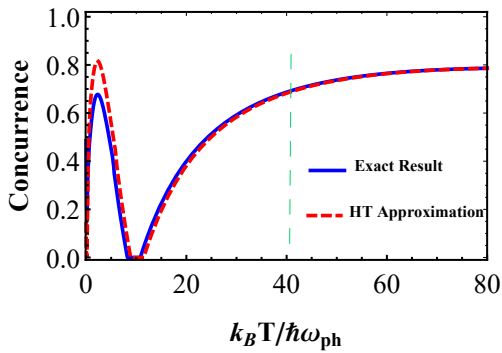}
\caption{Concurrence-Temperature behavior shows the characteristic temperature $k_{B}T\geq 40 \hbar\omega_{ph}$ that is due to the comparison between the exact result and high-temperature approximation for $\frac{eV}{\hbar\omega_{ph}}=0.1$ and $\kappa=0.63$.}\label{Fig5}
\end{figure}

For high temperature limit($k_{B}T \gg \hbar \omega_{ph}$), the phonon distribution function is approximated as $N_{ph}\simeq\frac{k_{B}T}{\hbar \omega_{ph}}$. In this case, the modified Bessel function for large argument is expressed as 
\begin{equation} 
I_{l}(z)\simeq\frac{e^z}{\sqrt{2\pi z}}(1-\frac{4l^{2}-1}{8z}+...), z\gg 1. 
\end{equation} 
Applying this expression into the equations (\ref{M+}) and (\ref{M-}) gives us: 
\begin{equation} 
M^{+,HT}_{\alpha \nu}= \Gamma_{\alpha \nu} \sum_{l} [1-f_{\nu}(-l\omega_{ph}+\epsilon_{\nu})] 
\frac{
e^{
\frac{-g_{ph}}{4}
\frac{\hbar \omega_{ph}}{k_{B}T}
}
}
{
\sqrt{
4\pi g_{ph}
\frac{k_{B}T}{\hbar \omega_{ph}}
}
}
\end{equation} 
and 
\begin{equation} 
M^{-,HT}_{\alpha \nu}= \Gamma_{\alpha \nu} \sum_{l} f_{\nu}(l\omega_{ph}+\epsilon_{\nu})
\frac{
e^{
\frac{-g_{ph}}{4}
\frac{\hbar \omega_{ph}}{k_{B}T}
}
} 
{
\sqrt{
4\pi g_{ph}
\frac{k_{B}T}{\hbar \omega_{ph}}
}
}. 
\end{equation} 
 Using these equations, we obtain the concurrence in the high-temperature limit. We present it in a comparison scheme between HT approximation and the exact result in Fig.(\ref{Fig5}). 
This figure introduces a characteristic temperature depending on the frequency of phonon. In this temperature which should be at least $40$ times larger than the phonon energy, concurrence in HT approximation can completely reproduce the exact result of entanglement. This analysis confirms that the entanglement in our proposed QDM model is strongly influenced by the phonon effects which in turn depends on the surrounding temperature. 

The main advantage of our proposed QDM setup is that the steady-state entanglement can be generated in a simple tunable junction. Also, the entanglement with the robust amount can be easily preserved by engineering the coupling of QDs-leads asymmetrically and manipulating the driven bias voltage.
In the future, it will be useful to investigate the stability of entanglement and entanglement revival in an array of quantum dot molecules under the effect of temperature dissipation to improve the building quantum computers.
\section{Conclusions}\label{Conclusion}
We studied the dynamics of entanglement formation in a carbon nanotube quantum dot molecule including phonon interactions in a biased junction. 
It was shown that the generated steady-entanglement between the coupled quantum dots could be controlled and preserved to achieve a robust value. This procedure was performed by implementing an asymmetric coupling strategy and also by applying a tunable external bias voltage. 
In response to an applied time-dependent bias voltage, the time evolution of entanglement exhibited periodic revival which could be reached the maximum magnitude. 
We could characterize the thermal entanglement degradation and revival phenomena through temperature evolution. In this case, the strength of phonon coupling influenced the rebirth of thermal entanglement comprehensively such that the revival occurred stronger for a higher phonon coupling amount. 
\section{Acknowledgments}
The authors wish to thank the Office of Graduate Studies and Research Vice President of the University of Isfahan for their support.
\section*{References}

\end{document}